\documentclass[%
twocolumn,%
showpacs,%
showkeys,%
preprintnumbers,%
amsmath,amssymb,%
page%
]{revtex4}
\usepackage{graphicx}
\usepackage{dcolumn}
\usepackage{bm}

\def\timeevol#1{\setbox1=\hbox{$\longmapsto$}\setbox2=\hbox{$
    \scriptstyle #1$}\copy1\kern-.5\wd1\kern-.5\wd2
    \raise-1.2\ht2\copy2\kern-.5\wd2\kern\wd1}

\begin{document}


\title{
Classical and Quantum Contents of Solvable Game Theory on Hilbert Space}
\author{
Taksu Cheon${ }^{1}$
and
Izumi Tsutsui${ }^{2}$
}

\affiliation{
${ }^1$
Laboratory of Physics, Kochi University of Technology,
Tosa Yamada, Kochi 782-8502, Japan\\
${ }^2$
Institute of Particle and Nuclear Studies,
KEK, Tsukuba, Ibaraki 305-0801, Japan
}

\date{April 28, 2005}

\begin{abstract}
A simple and general formulation of the quantum game theory is presented, 
accommodating all possible strategies in the Hilbert space for the first time.
The theory is solvable for the two strategy quantum game, which 
is shown to be equivalent to
a family of classical games supplemented by quantum interference.  
Our formulation gives a clear perspective to understand 
why and how quantum strategies outmaneuver classical strategies.  
It also reveals novel aspects of quantum games
such as 
the stone-scissor-paper phase sub-game and the fluctuation-induced moderation.
\end{abstract}


\maketitle

%
%
Attempts to extend game-theoretical strategies to the quantum domain \cite{ME99,EW99,MW00}
have attracted journalistic attention in the academic community and beyond, 
with an intriguing solution to the classical problem of
the Prisoner's Dilemma \cite{AX84}.
The substance of the emerging quantum game theory, however, 
is still shrouded in mystery,
and in spite of the rapid accumulation of literature \cite{IQ05}, 
we still find {\it ad hoc} assumptions and arbitrary procedures 
scattered in the field.
Quite naturally, there have been persistent doubts as to
their generality and finality \cite{BH01, EP02}.
%
{}For the quantum treatment of game strategies to become truly a theory, 
a workable framework to accommodate
all possible quantum states available in the system,
preferably with analytic solutions illuminating its structure, is highly desired.
In particular, it needs to clarify the reason behind 
the puzzling effectiveness of quantum strategies 
in situations where their classical counterparts fail
to give satisfactory results.
%
 
%
In this article, we attempt to answer this call with
a full Hilbert space formulation of the game theory.
It is shown that assigning vectors in a Hilbert space to game strategies
entails the introduction of an element that provides correlation for 
the strategies of the individual players.  
For two strategy games, the correlation is generated by operators that implement 
swapping and simultaneous renaming of the player's strategies.   
The quantum game is then split into two parts, one consisting of a family of classical games 
and the other representing the genuine quantum ingredient of the game.
The game, as a whole, is solvable.
We illustrate our formalism with numerical examples
on Prisoner's Dilemma and  discuss the classical and quantum contents
appearing in the Nash equilibria.  
We also point out the existence of such curious phenomena as
the stone-scissor-paper game found for phase variables of the strategy, 
and the quantum moderation which occurs for fluctuating correlations. 

%
%
%
To present our scheme of quantum game, we first consider 
Hilbert spaces ${\cal H}_A$ and ${\cal H}_B$ in which the strategies 
of the two players $A$ and $B$ are represented by vectors 
$ \left | \alpha \right>_A \in {\cal H}_A$ and $ \left | \beta \right>_B \in {\cal H}_B$. 
The entire space of strategy of the game is then given by the 
direct product ${\cal H} ={\cal H}_A \times {\cal H}_B$.  A vector in ${\cal H}$ 
represents a {\it joint strategy} of the two players 
and can be written as
\begin{eqnarray}
\label{jointst}
\left | \alpha, \beta; \gamma \right> 
= J(\gamma) \left | \alpha \right>_A \left | \beta \right>_B,
\end{eqnarray}
where the unitary operator $J(\gamma)$ provides quantum correlation 
({\it e.g.,} entanglement) for the separable 
states $ \left | \alpha \right>_A \left | \beta \right>_B$.
Note that $J(\gamma)$ is  independent of the players' choice and is 
determined by a third party, which is hereafter referred to as the {\it coordinator}. 

Once the joint strategy is specified with $J(\gamma)$, the players are to receive the payoffs, 
which are
furnished by the expectation values of self-adjoint operators $A$ and $B$:
\begin{eqnarray}
\label{QNash}
\Pi_A(\alpha, \beta; \gamma) 
&=& \left < \alpha, \beta; \gamma  | A  |  \alpha, \beta; \gamma  \right > ,
\\ \nonumber
\Pi_B(\alpha, \beta; \gamma) 
&=& \left < \alpha, \beta; \gamma  | B  |  \alpha, \beta; \gamma  \right > .
\end{eqnarray}
Each of the players then tries to optimize their strategy to gain the maximal payoff, 
and our question is to find, if any, a stable strategy vector which corresponds to
the quantum version of the Nash equilibrium.  
Namely, we seek a point  
$(\alpha, \beta)= (\alpha^\star, \beta^\star)$ in the strategy space 
at which the payoffs separately attain the maxima as
\begin{eqnarray}
\label{AlBeNash}
\left. \delta_\alpha \Pi_A (\alpha, \beta^\star; \gamma)\right|_{ \alpha^\star}  = 0,
\quad
\left. \delta_\beta \Pi_B (\alpha^\star, \beta; \gamma) \right|_{\beta^\star} = 0,
\end{eqnarray}
under arbitrary variations in $\alpha$ and $\beta$.

A symmetric quantum game is defined by requiring that the strategy spaces of 
the two players are the same 
in dimensionality, $\dim{{\cal H}_A}$ $=\dim{{\cal H}_B} = n$,
and that  the payoffs are symmetric for two players.
The latter condition is expressed as
\begin{eqnarray}
\label{symgame}
\Pi_A (\alpha, \beta; \gamma) = \Pi_B ( \beta, \alpha; \gamma)
\end{eqnarray}
in terms of identically labeled strategies for both players 
\begin{eqnarray}
\label{basis}
\left | \alpha \right>_A =  \sum_i \alpha_i \left | i \right>_A ,
\quad
\left | \beta \right>_B =     \sum_i \beta_i \left | i \right>_B ,
\end{eqnarray}
with complex numbers $\alpha_i$, $\beta_i$ 
normalized as $\sum_i \vert \alpha_i \vert^2 =  \sum_i \vert \beta_i \vert^2 = 1$.
Here we have used a  common  orthonormal basis for both of the players, 
namely, a set of strategies of the two players which are 
in one-to-one correspondence 
$\left| i \right>_A \leftrightarrow \left| i \right>_B$ for $i = 0, 1, \ldots, n-1$.
Introducing the swap operator $S$ by
\begin{eqnarray}
S \left | i , j \right> =  \left | j , i \right>
\end{eqnarray}
for the states $\left | i ,  j \right> = \left | i \right>_A \left | j \right>_B$,
we have $S \left | \alpha, \beta \right> = \left | \beta, \alpha \right>$ for general separable states 
$\left | \alpha, \beta \right> =  \left | \alpha \right>_A \left | \beta \right>_B$.   
For our convenience, we introduce two more operators $C$ and $T$ defined by
\begin{eqnarray}
C \left | i , j \right>  =  \left | {\bar i}, {\bar j} \right> ,
\quad
T \left | i , j \right>  =  \left | {\bar j}, {\bar i} \right> ,
\end{eqnarray}
where the bar represents the complimentary choice;
${\bar i} $ $=  (n-1)-i$.
The operator $C$ is the simultaneous renaming (conversion) of strategy for two players, 
and $T$ is the combination
$T = CS$.
These operators $\{S, C, T\}$ commute among themselves and 
satisfy $S^2=$ $C^2=$ $T^2 = I$, $T=SC$, $S=CT$ and $C=TS$. 
With the identity $I$, they form the dihedral group $D_2$.

In terms of  the correlated payoff operators,
%
\begin{eqnarray}
{\cal A}(\gamma) 
= J^\dagger(\gamma)A J(\gamma) , \quad  
{\cal B}(\gamma) 
= J^\dagger(\gamma)B J(\gamma) ,
\end{eqnarray}
we have
$\Pi_A (\alpha, \beta; \gamma)$
$=\left< \alpha,\beta \right| {{\cal A}(\gamma)} \left| \alpha,\beta \right> $.  
%
It is convenient to choose the unitary operator $J(\gamma)$ such that 
${\cal A}(0)$ is diagonal in the product basis $ \left | i , j \right>$.   The game is then symmetric
if ${\cal B}(0) = S {\cal A}(0) S$, in which case ${\cal B}(0)$ is diagonalized simultaneously 
with the eigenvalues swapped, 
%
\begin{eqnarray}
\left< i', j' \right| {\cal A}(0) \left| i, j \right>
\!\!\! &=& \!\!\!    A_{ij} \delta_{i' i}\delta_{j' j},
\\ \nonumber
\left< i', j' \right| {\cal B}(0) \left| i, j \right>
\!\!\! &=& \!\!\!    A_{ji} \delta_{i' i}\delta_{j' j}.  
\end{eqnarray}
Observe that
$\Pi_A (\alpha, \beta; 0) =$ $\Pi_B (\beta, \alpha; 0)=$ $\sum_{i,j} x_i  A_{ij} y_j$ 
with $x_i = \vert \alpha_i\vert^2$, $y_j = \vert \beta_j\vert^2$ being 
the probability of choosing the strategies 
$ \left | i \right>_A$, $ \left | j \right>_B$.  This means that, at $\gamma = 0$, 
our quantum game reduces to the classical game with the payoff matrix $A_{ij}$ 
under mixed strategies.  
%

%
%
Now we restrict ourselves to two strategy games $n = 2$.  
The entire Hilbert space is spanned by $\left| \alpha, \beta; \gamma \right>$ 
with the unitary operator $J(\gamma)$ in the form
\begin{eqnarray}
J(\gamma) =  J(0) \, e^{i \gamma_1 S / 2} e^{i \gamma_2 T / 2} ,
\end{eqnarray}
where $\gamma = (\gamma_1,  \gamma_2)$ are real parameters. 
Note that, on account of the relation $S+T-C=I$
valid for $n = 2$, only two operators are independent in the set $\{S, C, T\}$.
For simplicity, we assume that $A$ is diagonalized under the 
basis $\left | i , j \right>$, which implies $J(0) = I$ and ${\cal A}(0)=A$.
The correlated payoff operator $A(\gamma)$ is split into two terms 
\begin{eqnarray}
\label{Agam}
{\cal A}(\gamma) 
= {\cal A}^{\rm pc}(\gamma) + {\cal A}^{\rm in}(\gamma)
\end{eqnarray}
where ${\cal A}^{\rm pc}$ is the  ``pseudo classical'' term and 
${\cal A}^{\rm in}$ is the ``interference'' term given, respectively, by
\begin{eqnarray}
\label{Agam01}
{\cal A}^{\rm pc}(\gamma) 
\!\!\!&=&\!\!\! 
   \cos^2{\!\! {\gamma_1\over 2}} A   \!+\! 
 ( \cos^2{\!\! {\gamma_2\over 2}} \!-\! \cos^2{\!\! {\gamma_1\over 2}} ) S A S
  \!+\! \sin^2{\!\! {\gamma_2\over 2}} C A C ,
\nonumber \\
%
{\cal A}^{\rm in}(\gamma) 
\!\!\!&=&\!\!\! 
    { {i } \over {2} } \sin \gamma_1(AS \!-\! SA)
 + { {i} \over {2} }  \sin \gamma_2(AT \!-\! TA) .
\end{eqnarray}
Correspondingly, the full payoff is also split into two contributions from
${\cal A}^{\rm pc}$ and ${\cal A}^{\rm in}$ as $\Pi_A=$
$\Pi_A^{\rm pc}+ \Pi_A^{\rm in}$.   To evaluate the payoff, we may choose both
$\alpha_0$ and $\beta_0$ to be real without loss of  generality, 
and adopt the notaions $(\alpha_0, \alpha_1)=(a_0, a_1 e^{i\xi})$ 
and $(\beta_0, \beta_1)=(b_0, b_1 e^{i\chi})$.  The outcome is
\begin{eqnarray}
\label{payof0}
& &
\Pi_A^{\rm pc}(\alpha,\beta; \gamma)
=
\sum_{i,j} { a_i^2 b_j^2  {\cal A}^{\rm pc}_{i j} }(\gamma) ,
\\ \nonumber
%
& &
\Pi_A^{\rm in}(\alpha,\beta; \gamma)
\\ \nonumber
& & 
= - a_0 a_1 b_0 b_1 
[ G_+(\gamma) \sin(\xi + \chi) + G_-(\gamma)  \sin(\xi - \chi) ] ,
\quad
\end{eqnarray}
with 
${\cal A}^{\rm pc}_{i j} (\gamma) = \left< i, j \right| {\cal A}^{\rm pc}(\gamma)  \left| i, j \right>$ 
and 
\begin{eqnarray}
\label {FGH}
G_+(\gamma)  \!\!\!& = &\!\!\!  (A_{00}  -  A_{11}) \sin\gamma_2,
\\ \nonumber 
G_-(\gamma)   \!\!\!& = &\!\!\!  (A_{01}  -  A_{10})\sin\gamma_1.
\end{eqnarray}
The split of the payoff shows that  the quantum game consists of two ingredients.  
The first is the pseudo classical ingredient associated with $\Pi_A^{\rm pc}$ in (\ref{payof0}), 
whose form indicates that we are, in effect, 
simultaneously playing three different classical games, {\it i.e.,} the original classical
game $A$, the altruistic game $SAS$ \cite{CH03} and the converted-value game $CAC$
with the mixture
specified by given $\gamma_1$ and $\gamma_2$.   
Regarding $\gamma$ as tunable parameters, we see that the quantum game 
contains a {\it family} of classical games that includes the original game.  
The second ingredient of the quantum game is the purely quantum component  $\Pi_A^{\rm in}$, 
which occurs only when both of the two players adopt quantum strategies 
with $a_0 a_1 b_0 b_1 \ne 0$ and non-vanishing phases $\xi$ and $\chi$.   
The structure of $\Pi_A^{\rm in}$ suggests that this interference term cannot be simulated by 
a classical game and hence represents the {\it bona fide} quantum aspect.
%
%

%
%
\begin{table}
\caption{
The quantum Nash equilibria with edge strategies.  The conditions for their appearance and 
their maximal payoffs under variations of $\gamma$ are shown.
}
\begin{ruledtabular}
\begin{tabular}{ccccc}
$\left| \alpha^\star,\beta^\star\right>$ &
 $\left| 0,0\right>$ & $\left| 1,1\right>$ & $\left| 0,1\right>$ & $\left| 1,0\right>$ \\
\hline
Condition &
  $\ H_+ > 0 $ & $\ H_- > 0 $ & $H_+ < 0$ & $H_+ < 0$ \\ 
&   &   & $H_- < 0$ & $H_- < 0$ \\
\hline
Max($\Pi_A^\star$) & 
  $A_{00}$ & $A_{00}$  & $A_{11}$ & $A_{01}\!+\!A_{10}\!-\!A_{11}$  \\
Max($\Pi_B^\star$) &
  $A_{00}$ & $A_{00}$ &  $A_{01}\!+\!A_{10}\!-\!A_{11}$ & $A_{11}$  \\
\end{tabular}
\end{ruledtabular}
\end{table}
We can find the quantum Nash equilibrium strategy explicitly by considering the condition 
(\ref{AlBeNash}) for the payoff obtained above.    
It can be readily confirmed that, modulo arbitrary phases,  the ``edge'' strategies  
$\left| \alpha^\star, \beta^\star \right>$  given by
\begin{eqnarray}
\label {tNE}
\left| \alpha^\star, \beta^\star \right> 
= \left| 0, 0\right>, \,  \left|1, 1\right>, \, \left|0, 1\right>, \, \left|1, 0\right>
\end{eqnarray}
can furnish Nash equilibria, depending on the  
signs of the functions
\begin{eqnarray}
\label {convtool}
H_\pm(\gamma) =  \tau(A) \pm \left[ G'_+(\gamma) + G'_-(\gamma) \right] ,
\end{eqnarray}
where  we define $\tau(A)=A_{00}\!-\!A_{01}\!-\!A_{10}\!+\!A_{11}$ 
and also
$G_+'(\gamma) = (A_{00} \!-\!  A_{11}) \cos\gamma_2$,
$G_-'(\gamma) = (A_{01}  \!-\!  A_{10})\cos\gamma_1$.
%
The precise conditions for the appearance of the equilibria, together with their maximal payoffs 
$\Pi_A^\star(\gamma) = \Pi_A(\alpha^\star(\gamma),\beta^\star(\gamma);\gamma)$
obtained under variations of $\gamma$, are summarized in TABLE  I.

%
%
\begin{figure*}
\includegraphics[width=4.4cm]{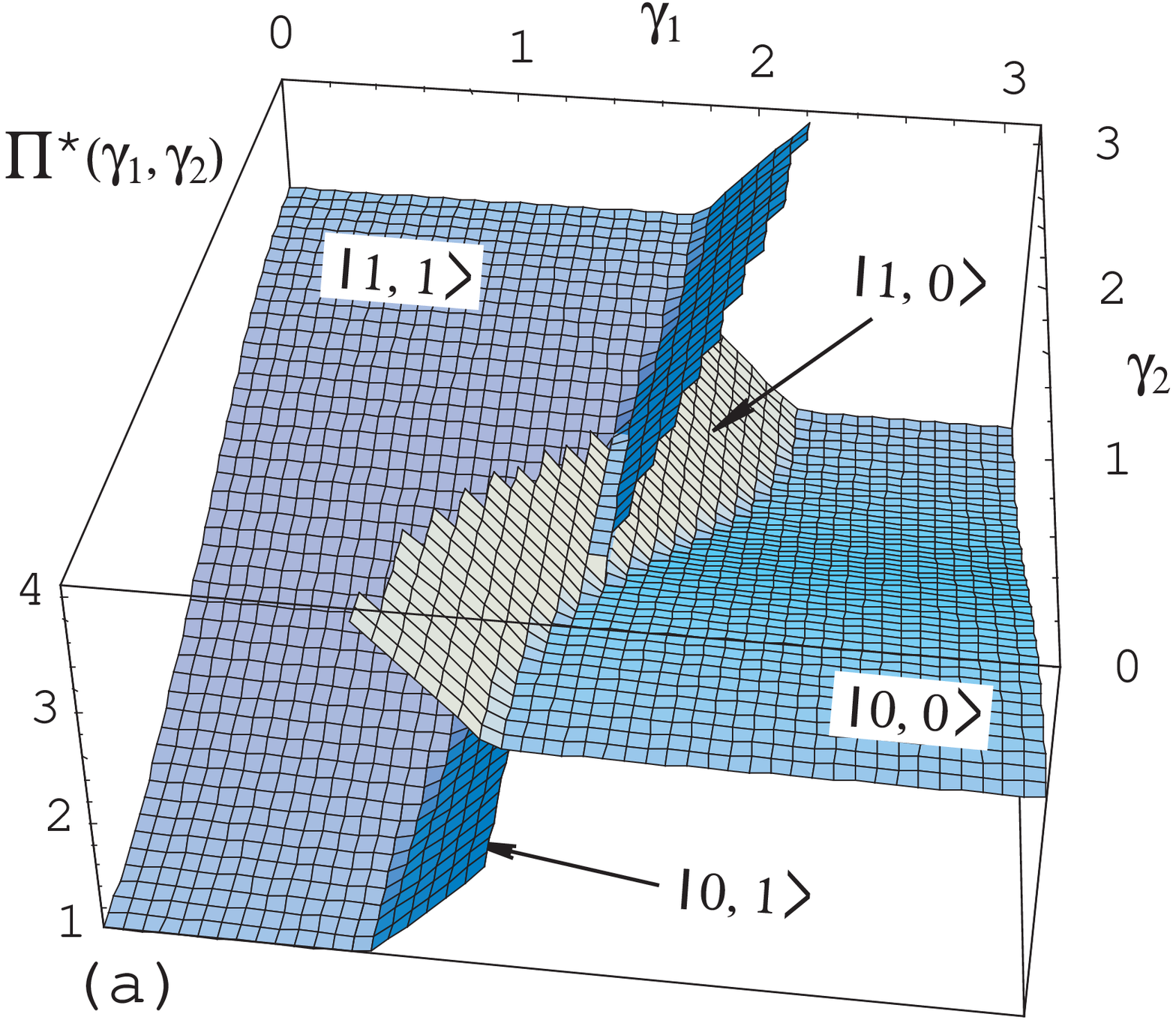} \quad\quad\quad
\includegraphics[width=4.4cm]{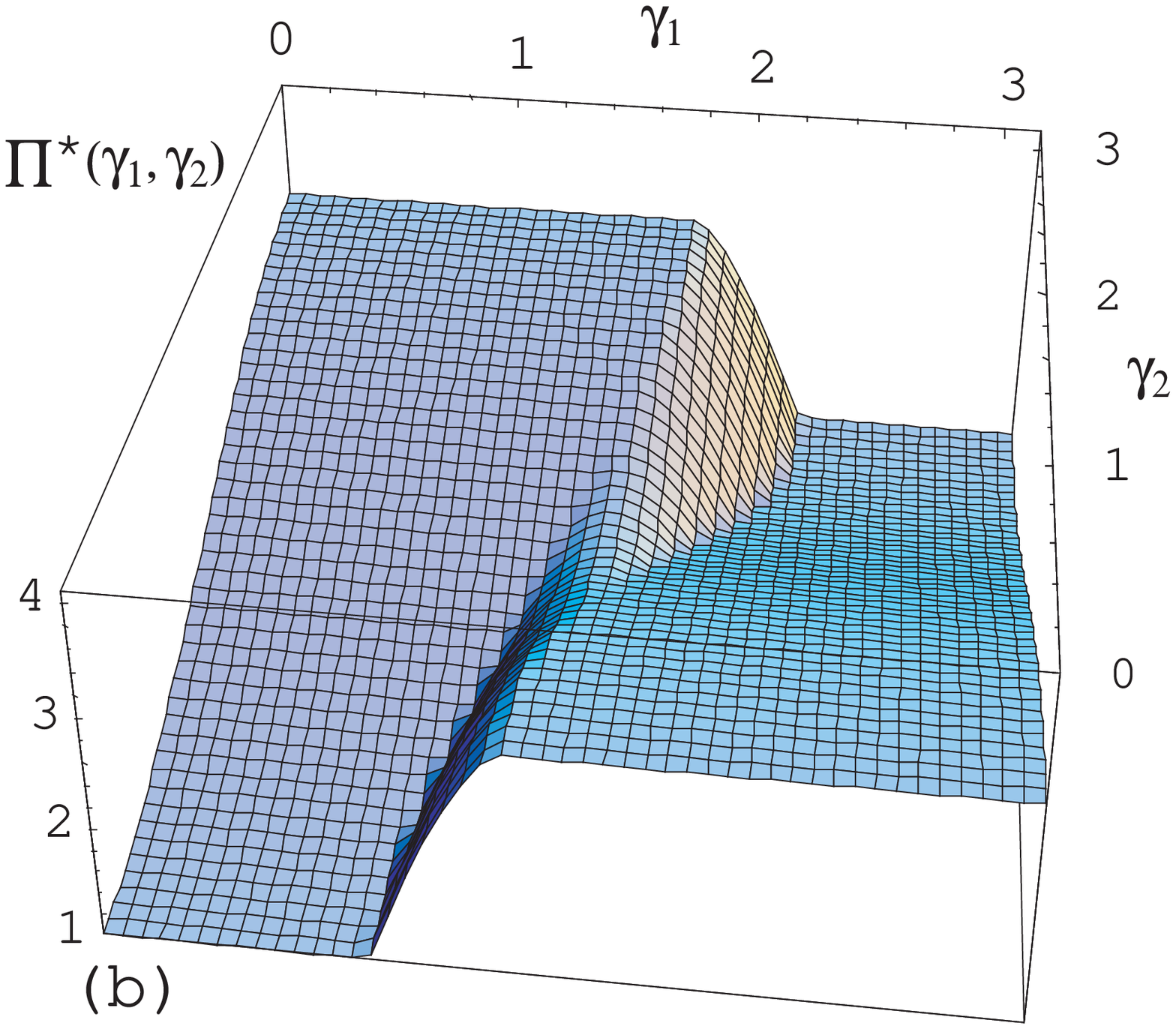} \quad\quad\quad
\includegraphics[width=4.4cm]{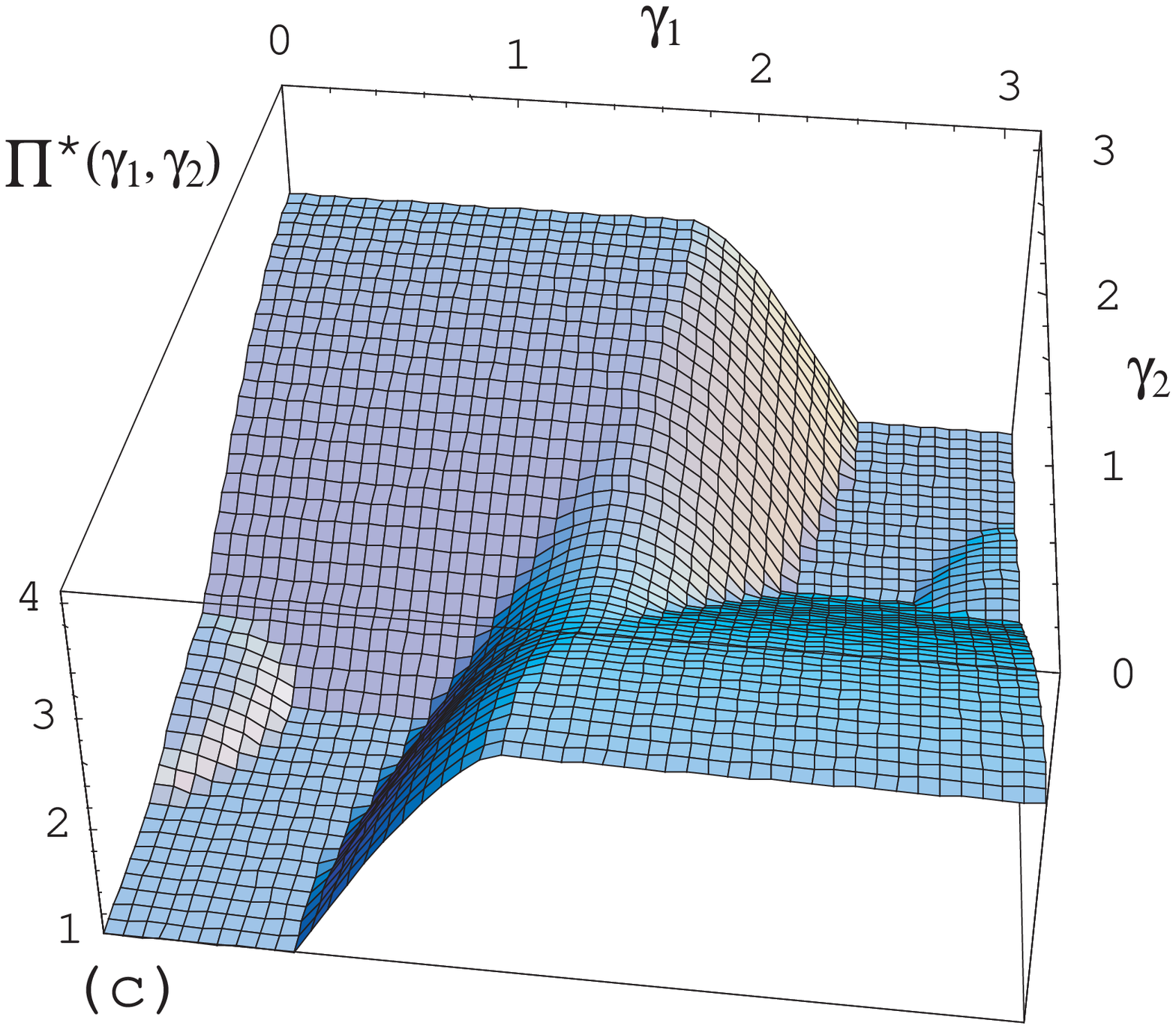}
\caption
{\label{fig1}
The quantum Nash equilibrium  payoff
$\Pi_A^\star(\gamma) = \Pi_A(\alpha^\star,\beta^\star;\gamma)$ 
(or ${\Pi}_A(P_A^*, P_B^*; \gamma)$ for mixed quantum strategies) 
as a function of $\gamma$.  
Only the region $\gamma_1, \gamma_2 \in [0,\pi]$  is shown since  
$\Pi_A^\star(\gamma)$ has the reflection invariance 
$\Pi_A^\star(2\pi-\gamma_1,\gamma_2)$
$= \Pi_A^\star(\gamma_1, 2\pi-\gamma_2)$$= \Pi_A^\star(\gamma_1,\gamma_2)$.
The extra invariance $\Pi_A^\star(\pi-\gamma_1, \pi-\gamma_2)$
$= \Pi_A^\star(\gamma_1,\gamma_2)$ is also visible.
(a)  Edge state Nash equilibria for $A_{00} \!=\! 3$, $A_{01} \!=\! 0$, $A_{10} \!=\!5$ 
and $A_{11} \!=\! 1$.  The  value  $(\gamma_1,\gamma_2) \!=\! (0.9272, 0)$
gives the maximum payoff $\Pi_A^\star=4$ for one of the players.
(b)  Symmetric mixed quantum Nash equilibrium with the same parameters as (a).   
The maximum payoff $\Pi_A^\star=3$ is obtained at $\gamma_2^\star = 0$ 
and $1.3694 \le \gamma_1^\star< \pi$.
(c)  Mixed Nash equilibrium for $A_{00} \!=\! 3$, $A_{01} \!=\! 0$, $A_{10} \!=\! 5$ 
and $A_{11} \!=\! 0.2$.
The two bumps near the left and right ends are due the pure symmetric Nash equilibria (\ref{MiNa}).
}
\end{figure*}
%
%
{}For illustration, let us consider the case where the classical game 
$\gamma = 0$ exhibits the Prisoner's Dilemma, $A_{01} <$$A_{11} <$$ A_{00} <$$ A_{10}$.  
Adopting the numerical values used in  \cite{EW99},  we observe from 
FIG. 1(a) that two asymmetric Nash equilibria coexist  
in the middle strip that separates the two domains where
the symmetric Nash equilibria arise.
The maximal payoff for player $A$ is achieved by the equilibrium strategy $\left| 1,0 \right>$ 
at the optimal choice $(\gamma_1^\star, \gamma_2^\star)$$=( 2\arcsin\sqrt{\lambda}, 0)$,
and also by its symmetric partner  $\left| 0,1 \right>$ at $( \pi-2\arcsin\sqrt{\lambda}, \pi)$
where we use
\begin{eqnarray}
\label {optchoice}
\lambda = (A_{11} - A_{01})/(A_{10} - A_{01}).
\end{eqnarray}
%
Interestingly, the maximal payoff is better  
than that obtained when the players decide to ``deny'' $\left| 1, 1\right>$ 
or ``confess'' $\left| 0, 0\right>$.  
Note that the joint strategies (\ref{jointst}) of the players realized at these Nash equilibria are
actually entangled due to the correlation factor $J(\gamma)$.   Indeed,  
the entropy of entanglement $S(\rho_{\rm red})$ evaluated for the reduced density operator 
$\rho_{\rm red}$ \cite{BB96} of the optimal state reads 
%
\begin{eqnarray}
\label {entent}
S(\rho_{\rm red}) = - \lambda \log \lambda - (1-\lambda) \log (1-\lambda),
\end{eqnarray}
which is nonvanishing since  $0 < \lambda < 1$ for the Prisoner's Dilemma.  
The optimal equilibrium, however, does not provide a desired resolution for the dilemma, 
because it is achieved at the expense of player $B$ receiving a lower payoff.   
In fact, in the middle strip,
the original Prisoner's Dilemma
turns into the Game of Chicken which has its own dilemma of a different kind.    
%

%
Leaving the numerical example aside for now,
we return to the general case, and examine 
the possibility of a pure Nash equilibrium which is not one of 
the edge states. 
The interference term $\Pi_A^{\rm in}$ now comes into play, and
applying the condition for phases,
$\delta_\xi \Pi_A |_{\xi=\chi=\xi^\star}$, we obtain
\begin{eqnarray}
\label{PhNa}
\cos 2\xi^\star = - {G_-(\gamma)} / {G_+(\gamma)} .  
\end{eqnarray}
When $| G_- | \le | G_+ |$,
there is an equilibrium solution $\xi=\chi=\xi^\star$ with the payoff
$\Pi_A^{\rm in} = a_0 a_1 b_1 b_1\Delta(\gamma)$, where 
$\Delta(\gamma)=\sqrt{G_+^2-G_-^2}$.
The condition for the amplitudes $\delta_{a_0} \Pi_A = 0$ 
and $\delta_{b_0} \Pi_B = 0$ then provides, 
along with the edge state solutions (\ref{tNE}),
the symmetric solution,
\begin{eqnarray}
\label{MiNa}
a_0^\star = b_0^\star = 
\left(
{ { H_-(\gamma) - \Delta(\gamma) } \over 
   { H_+(\gamma) + H_-(\gamma) -  2 \Delta(\gamma) } } 
\right) ^{1/2} ,
\end{eqnarray}
which is valid ($0\le a_0^\star \le 1$) if  $(H_+  - \Delta)(H_-  - \Delta) \ge 0$.   
There is no asymmetric pure Nash equilibria apart from the two edge solutions.

When $ | G_- | > | G_+ |$, 
there is no dominant strategy for the phases:
the player $A$ tries to top player $B$ by choosing a phase which is off by $\pi/2$
to maximize $\Pi_A^{\rm in}$.
The player $B$ does the same, and if the game is played repeatedly, the result is
a uniform random distribution for both $\xi$ and $\chi$.
This is a continuous version
of the paper-scissor-stone game for phases, 
and results in the zero average for the interference term, $\Pi_A^{\rm in} = 0$.   
Thus, we reach formally the same symmetric solution (\ref{MiNa})
with $\Delta(\gamma) = 0$.
The existence requirement of the solution simplifies to $H_+ H_-  \ge 0$ for this case, 
and we find, from TABLE I, that the equilibrium appears precisely 
in the region of the asymmetric pure Nash equilibria.    
The foregoing argument is made formal by 
considering the {\it mixed} quantum strategies
specified by
probability distributions 
$P_A(\alpha)$, $P_B(\beta)$ over the players' actions normalized 
under suitable measures $d\alpha$, $d\beta$.
The distribution-averaged payoff can be defined as
%
$
{\Pi}_X(P_A,P_B; \gamma)
= \int d\alpha d\beta  P_A(\alpha) \Pi_X(\alpha, \beta; \gamma) P_B(\beta)
$
%
for $X=A, B$.  The players seek a distribution $P_A=P_B=P^\star$
which simultaneously maximize ${\Pi}_A(P_A,P_B; \gamma)$ 
and ${\Pi}_B(P_A,P_B; \gamma)$.
Such a distribution furnishes a mixed quantum Nash equilibrium, extending
the concept of
the pure quantum Nash equilibrium specified by single values of $\alpha$ and $\beta$.
Note, however, that the latter is already probabilistic in terms of classical strategies, 
possessing a classical mixed strategy game as a subset.  
The former, on the other hand,  is probabilistic in terms of quantum strategies, 
and is realized by an ensemble of quantum systems.  
%
%
%
%

%
In FIG. 1(b), the mixed quantum Nash equilibrium payoff 
for the Prisoner's Dilemma of FIG. 1(a) is shown as a function of $\gamma$.
In the middle strip, asymmetric equilibria, 
which are known to be dynamically unstable  \cite{WE95}, 
are now replaced by the mixed
Nash equilibrium given by (\ref{MiNa}) with $\Delta(\gamma)=0$.  
The global maximum of the payoff $\Pi_A^\star(\gamma^\star)=A_{00}$ is attained along the line 
$\gamma_2^\star=0$, $2\arcsin\sqrt{\eta} \le \gamma_1^\star \le \pi$ 
with $\eta =(A_{10}\!-\!A_{00})/(A_{10}\!-\!A_{01})$.
We mention that the quantum Nash equilibrium found in \cite{EW99} 
corresponds to $(\gamma_1, \gamma_2)=$ $(\pi/2, 0)$ in our scheme.   
Unlike the optimal point $\left| 1,0 \right>$,  
the joint strategy state of this equilibrium remains to be $\left| 0,0 \right>$ and hence 
is {\it not} entangled.  
In fact, the entanglement of the Nash equilibrium is inessential in this example,
since  the interference term vanishes for all parameter values, leaving only
classically interpretable terms. 
The truly quantum characteristics of the game manifests itself 
when the non-edge solution (\ref{MiNa}) appears.
We show an example of 
such cases in FIG. 1(c) which is obtained by modifying 
the payoff parameters slightly from the previous ones.  
Here, the Nash equilibria (\ref{MiNa}) 
contributes to the increase of the payoff as seen by the 
convex structures at the two ends.
Examination of the solution to other types of quantum games \cite{FA03,OZ04a}
would naturally be the next task.
%
%
%
%

There are several different interpretations possible for the role of the
coordinator $J(\gamma)$.
The first is that it furnishes the unitary family of payoff operators ${\cal A}(\gamma)$ 
from a given classical payoff matrix $A_{ij}$, and 
therefore acts independently from the two players.
The second is that it acts as a collaborator to the players and serves to maximize 
the payoff at the Nash equilibria by tuning $\gamma$ as
$\delta_{\gamma} \Pi_A|_{\gamma^\star} = 0$.  
In the above numerical examples, we have started from the first interpretation 
and tacitly moved to the second.
Yet another interpretation of the coordinator is that it 
generates quantum fluctuations for the payoffs by randomizing the parameter $\gamma$.   
In this case, the game is effectively given as an average over the fluctuations, 
and may be studied by integrating out the parameters $\gamma_1$ and $\gamma_2$.  
At the level of the payoff operator, the outcome is expressed as
\begin{eqnarray}
\label{avapayoff}
\int d\gamma_1 d\gamma_2 \, {\cal A} (\gamma)  = {1 \over 2} A + {1 \over 2} CAC,
\end{eqnarray}
which implies that the quantum fluctuations yield {\it quantum moderation} to 
the game by washing out the individual's preference.


%

%
We conclude this article by examining the quantum
and classical aspects of strategies in our formulation.
We first stress that our treatment is based on the full set of quantum strategies, 
and thus the quantum Nash equilibria obtained here are truly optimal within the entire Hilbert space.

Among the quantum Nash equilibria, those obtained within
the classical family can always be simulated by some classical means, even when 
their joint strategy states are entangled in the Hilbert space.
In a sense, these classically realizable quantum strategies are 
the possible link between the quantum game theory 
and the classical games in macroscopic social and ecological settings. 
In contrast, the quantum solutions that arise only with the interference terms
represent the first genuinely quantum Nash equilibria,
offering superior payoffs, that have no counterparts in classical strategies,

{}For the comprehensive classification of the quantum games in our scheme,  
the full analysis of the classical family is indispensable.
This should also pave the way to the extension for
games with more than two strategies and two players. 
\\

One of the authors (TC) thanks the members of the
Institute of Particle and Nuclear Studies at KEK for the hospitality granted to him
during his entended stay.


\end{document}